\documentclass[12pt]{article}
\usepackage{amsmath}
\usepackage[dvips]{epsfig}
\usepackage{amssymb}

\def\R{\mathbb R}
\def\C{\mathbb C}

\newtheorem{theorem}{Theorem}[section]
\newtheorem{lemma}[theorem]{Lemma}

\begin{document}

\title{Multipeakons and the Classical Moment Problem}

\author{Richard Beals\thanks{Research supported by NSF grant DMS-9800605}\\
Yale University \\New Haven, Connecticut
\and
David H. Sattinger\thanks{Research supported by NSF Grant DMS-9971249. Present address: Head, Department of Mathematics and Statistics, Utah State University, Logan, Utah, 84322}\\
University of Minnesota\\Minneapolis, Minnesota
\and
Jacek Szmigielski\thanks{Research supported by the Natural Sciences and 
Engineering Research Council of Canada}\\
University of Saskatchewan\\
Saskatoon, Saskatchewan}

\maketitle

\centerline{to appear: {\it Advances in Mathematics}}

\medskip

\centerline{\it Dedicated to Professor Gian-Carlo Rota}

\medskip

{\abstract Classical results of Stieltjes are used to obtain explicit 
formulas for the peakon-antipeakon solutions of the Camassa-Holm equation. The closed form solution is expressed in terms of the orthogonal polynomials of the related classical moment problem.
It is shown that collisions occur only in peakon-antipeakon pairs, and the details of the collisions are analyzed using results {}from the moment problem. A sharp result on the steepening of the slope at the time of collision is given. Asymptotic formulas are given, and the scattering shifts are calculated explicitly.}

\medskip

\medskip

\section{Introduction}

The Korteweg-deVries equation is a simple mathematical model for gravity waves
in water, but it fails to model such fundamental physical phenomena as the 
extreme wave of Stokes \cite{stokes}. The failure of weakly nonlinear dispersive equations, such 
as the Korteweg-deVries equation, to model the observed breakdown of regularity
in nature, is a prime motivation in the search for alternative models for nonlinear dispersive waves \cite{rosenau}, \cite{whitham}. 

In 1976 Green and Naghdi \cite{gn} derived a system of water wave equations to model fluid flows in thin domains, such as internal waves in coastal regions. The Green-Naghdi equations have a Hamiltonian structure, and in 1993 Camassa and Holm \cite{ch} used scaling and an asymptotic expansion to obtain, in the one dimensional case, an approximate Hamiltonian which is formally integrable by the method of inverse scattering. The strongly nonlinear equation they obtained,
\begin{equation}\label{chequ}
 u_t-\frac14 u_{xxt}+\frac32 (u^2)_x-\frac18 (u_x^2)_x-
\frac14 (uu_{xx})_x=0, 
\end{equation}
supports solutions, dubbed ``peakons'', that are continuous but only piecewise analytic. Equation \eqref{chequ} had originally been
obtained by B. Fuchssteiner \cite{fuch} by the method of recursion operators in 1981. He showed the equation was Hamiltonian, but gave no physical interpretation, nor an isospectral operator; and the equation attracted no special attention until its rediscovery by Camassa and Holm.

Motivated by the form of traveling wave solutions of \eqref{chequ}, Camassa and Holm proposed solutions of the form (in the normalization used in this paper)
\begin{equation}\label{npeakon}
u(x,t)=\frac12\sum_{j=1}^n m_j(t)\exp(-2|x-x_j(t)|).
\end{equation} 
to represent $n$ interacting traveling waves.  They substituted this 
{\it Ansatz\/} into \eqref{chequ} and obtained a Hamiltonian 
system of equations for $m_j$, $x_j$; the Hamiltonian is obtained directly 
by substituting \eqref{npeakon} into the formal Hamiltonian for \eqref{chequ}.
This system describes geodesic flow on a manifold with metric tensor 
$g^{ij}=\exp(-2|x_j-x_j|)$. A solution \eqref{npeakon} can contain
both peaks $m_j>0$ and antipeaks $m_j<0$; at large positive or negative
time it is asymptotic to a superposition of single travelling
waves: peakons and antipeakons.  

In \cite{chh} the Hamiltonian system for two peakons is integrated, and explicit expressions for the 
relative position $x_1-x_2$ and relative momentum $p_1-p_2$ are obtained.
A qualitative analysis of the interaction of 
peakons, and of the collision of a perfectly antisymmetric peakon-antipeakon pair, together with a number of numerical studies are carried out, and a formula for the phase shifts of the interaction of two solitons was obtained.

\medskip

\begin{figure}[!ht]
\centerline{\epsfxsize=\linewidth \epsfbox{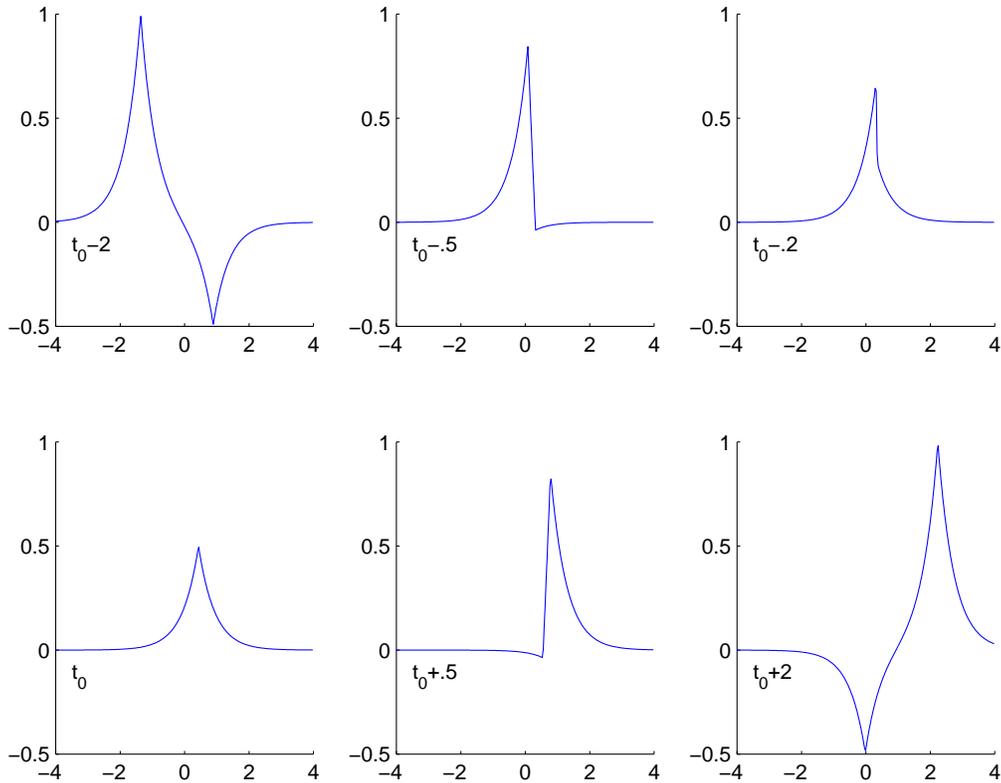}}
\caption{ A peakon moving {}from left to right collides at time 
$t_0=.2310$ with an antipeakon moving right to left, 
computed {}from the explicit formulas using Matlab. 
Here $\lambda_-=-1$, $\lambda_+=2$, 
$a_\pm (0)=.5$. The slope becomes infinite at the instant of collision. 
Sharp results on the steepening of the slope are given in
\S\ref{collisions}. The vertical scale is exaggerated; the peakons are n
ot as sharp as they appear here.}
\end{figure}

\medskip

The strongly nonlinear dispersive wave equation \eqref{chequ}
possesses a remarkable mathematical structure and is the subject of a 
steadily growing literature. 
In \cite{bss2} we used the theory of continued fractions
and formulas of Stieltjes \cite{stieltjes2} (based on prior results of 
Frobenius \cite{frobenius}) to give  algebraic formulas 
for the multipeakon solutions (all $m_j>0$).  These formulas led to explicit expressions for the asymptotic behavior of the positions $\{x_j\}$ and momenta $\{m_j\}$ as $t\to\pm\infty$, and thus to explicit calculation of the scattering shifts undergone by the peaks as a result
of their interaction.  

\medskip

In this paper we give a brief but self-contained discussion of the
spectral problem associated with \eqref{chequ}, detailed proofs 
of the results announced
in \cite{bss2}, and an extension of those results to the 
case of peakon/antipeakon interactions.  In the remainder of this
section we outline the paper, place it in the context
of related work, and describe the conclusions concerning
the solutions \eqref{npeakon}.

In \S\ref{chev} we obtain the multipeakon/antipeakon 
solutions as a restriction of 
\eqref{chequ} to a singular subclass. This necessitates a 
reinterpretation of the Lax pair in the sense of distributions. In 
\S\ref{liouville} we describe the Liouville transformation that transforms 
the spectral problem on the line to a formal density problem for a finite 
string, first investigated by M. G. Krein. In \S\ref{weylfct} we describe the 
Weyl function, which encodes the spectral data, and its expansion in continued
fractions. In \S\ref{stieltjes} we describe the formulas 
by which the continued fraction expansion is recovered {}from the 
Laurent series for the Weyl function.  In \S\ref{asymp} we use the
explicit formulas to obtain the 
asymptotic behavior of the multipeakon solutions and the scattering shifts 
of the peakons and antipeakons that result {}from their interactions. 

A closed form of the multipeakon/antipeakon solution is given (Theorem \ref{momform}) in terms of the orthogonal polynomials associated with the moment problem. 

Stieltjes' original theory was restricted to positive weights, but the formulas are algebraic and extend to weights of mixed sign, so long as certain determinants do not vanish.  In the present context, these determinants are functions of time.  Collisions of peakons and antipeakons occur precisely when one or more of these determinants vanish.  At such a point some of the weights $m_j$ become infinite, but the solution itself remains bounded throughout the 
collision. Moreover, there can be no triple collisions:
at a given time, collisions occur only in distinct peakon-antipeakon
pairs: a peakon moving to the right encounters and passes an antipeakon moving to the left.  We prove these 
facts in \S\ref{collisions}, using the explicit solutions together with 
the oscillation properties of the orthogonal polynomials of the associated moment problem.

A primary reason for the interest in \eqref{chequ}
is that it is an integrable model for the breakdown of regularity, a phenomenon not modeled by the Korteweg-deVries equation. A number of qualitative results of a general nature on the steepening of the slope at the instant of breakdown have been obtained by analytic methods \cite{chh}, \cite{conesch1},
\cite{conesch2}; McKean,\cite{mckean2}, has given a sharp result for the general case.  The exact results obtained {}from the Stieltjes formulas yield the same result (Theorem \ref{slope}) concerning the steepening of the slope as the instant of collision of a peakon and antipeakon is approached.

\medskip

In \cite{moser}, \cite{moser2}, J. Moser applied the theory of continued 
fractions to finite Jacobi ({\it a.k.a.\/} Toda) flows. He obtained an 
explicit solution for $2\times 2$ and $3\times 3$ Jacobi matrices, though he  did not use the full results of Stieltjes. He nevertheless deduced qualitative properties of the solution to the isospectral deformations of finite Jacobi matrices, and calculated the scattering shifts. The relationship of Moser's work, as well as the classical formulas of Frobenius and Stieltjes, to the Riemann theta functions is discussed in an unpublished note of H. P. McKean \cite{mckean}.

Alber {\it et al.\/} \cite{alber}, \cite{alber2} obtained a number of special solutions of the Camassa-Holm equation, including peakons, $n-$solitons, and quasiperiodic solutions, by inverting the transformation to action-angle variables. They attacked the Hamiltonian systems directly, without the use of inverse scattering theory.  They assert that the general inversion problem can  be solved in terms of Riemann theta functions; but explicit results, such as the phase shift undergone by interacting solitons, are obtained only in the case $n=2$. 

Moser points out that the moment problem was the forerunner of modern spectral theory.  Boris Levitan once remarked to one of the present authors that it was Gelfand's interest in a continuum analogue of the classical moment problem that led to their celebrated solution of the inverse spectral problem. B. Simon in \cite{bs} appears to close the circle, and returns to the original view of inverse spectral theory as a continuous analogue of the moment problem.

The theorem of Stieltjes and the method of continued fractions was first 
applied to an inverse problem for ordinary differential equations by M. G. Krein \cite{krein1}, \cite{krein2}. The method was extended to a 
special class of fourth order equations by V. Barcilon \cite{ba4}, 
which suggests that there are interesting isospectral deformations of such fourth order equations. 

The multipeakon/anti\-pea\-kon solutions are intimately related to Jacobi matrices, but differ {}from the Toda flows in a fundamental way.  The Toda flow describes the dynamics of a lattice of constant weights joined by restoring forces; the conjugate variables are the relative positions of the masses, and their momenta.  In the multipeakon flow, the weights $m_j$ vary in time and may take on both positive and negative values; the conjugate variables are the positions of the weights, and the weights themselves. We plan to investigate the relationship of the Jacobi flows to the moment problem in a future paper.

\section{The Camassa-Holm equation}\label{chev}

Equation \eqref{chequ} is obtained as the compatibility condition for
an overdetermined system \cite{ch}; in the present normalization we take
\begin{gather}
L_0(z)f=0, \qquad \frac{\partial f}{\partial t}=A_0(z)f \label{lxpair}\\[4mm]
L_0(z)=D^2-z\,m(x,t)-1,\qquad 
A_0(z)=\left(\frac1z-u(x,t)\right)D+\frac12 u_x(x,t)\nonumber 
\end{gather}
where $x\in \R,\ z\in\C$.

Differentiating the first equation in \eqref{lxpair} with respect to
$t$  and the second twice with respect to $x$ and setting $f_{xxt}=f_{txx}$,  
we obtain, after some calculations,
\begin{equation}
m_t+(um)_x+m\,u_x=0,
\qquad
2m=4u-u_{xx}.\label{ch1}
\end{equation}

We assume here that $m$ and its derivatives vanish at $\pm \infty$.
According to the second equation in \eqref{ch1} the same is then true of $u$ and $u_x$.

Conversely, suppose that the function $m$ evolves according to \eqref{ch1},
with $m$ and its derivatives vanishing at $x=\pm\infty$.
This is also the compatibility condition for two modifications of
\eqref{lxpair}:
\begin{gather}
L_0(z)\varphi_0=0, \qquad \frac{\partial \varphi_0}{\partial t}=
\Big(A_0(z)-\frac1z\Big)\varphi_0 \label{lxpair1}\\[4mm]
L_0(z)\psi_0=0,\qquad \frac{\partial \psi_0}{\partial t}=
\Big(A_0(z)+\frac1z\Big)\psi_0. \label{lxpair2}
\end{gather}
These evolution equations are consistent with the asymptotic
conditions
\begin{eqnarray}
 \varphi_0(x,z)&\sim e^{x},\quad x\to-\infty;\label{wave1}\\
\psi_0(x,z)&\sim e^{-x},\quad x\to+\infty,\label{wave2}
\end{eqnarray}
respectively.
It follows that there are unique {\it wave functions\/} $\varphi_0(x,t,z)$,
$\psi_0(x,t,z)$ that satisfy 
\eqref{lxpair1}, \eqref{wave1}, and \eqref{lxpair2}, \eqref{wave2},
respectively.

The remaining $x$--asymptotics of $\varphi_0$ are necessarily
\begin{equation}
\varphi_0(x,t,z)\sim b(t,z)\,e^{x}+c(t,z)\,e^{-x}, 
\qquad
x\to + \infty.\label{plusinf}
\end{equation}
It follows {}from the time evolution of $\varphi_0$ that
\begin{equation}\label{dataev}
\frac{\partial b}{\partial t} =0, \qquad \frac{\partial c}{\partial t}
=-\frac{2c}z.
\end{equation}
The eigenfunctions for this spectral problem are, by definition, the
functions $\varphi$ for those values $z=\lambda_\nu$
for which $b(\lambda_\nu)=0$.  Therefore \eqref{dataev} implies that the spectrum
$\{\lambda_\nu\}$ is invariant under the flow.  The coupling constants
are the values $c_\nu(t)=c(t,\lambda_\nu)$; they are characterized by
the relation
\begin{equation}\label{couple1}
\varphi_0(x,\lambda_\nu)=c_\nu\psi_0(x,\lambda_\nu).
\end{equation}
The evolution is given by specializing \eqref{dataev}:
\begin{equation}\label{coupleev}
\dot c_\nu=-\frac{2c_\nu}{\lambda_\nu}.
\end{equation}

Thus far we have tacitly assumed that $m$ is a continuous density.
The interesting case for present purposes is when $m$ is taken to be a discrete measure with weights $m_j$ at locations $x_j$:
\begin{equation}\label{discrete1}
m=\sum^n_{j=1}m_j\delta_{x_j},\qquad x_1<x_2<\dots< x_n.
\end{equation}
The two equations in \eqref{lxpair} are readily interpreted in the
sense of distributions.  As we note below, the function $u$ of
\eqref{ch1} arises in the same way.  It can be calculated explicitly
{}from \eqref{discrete1} and the 
second equation in \eqref{ch1}. In the normalization of the spectral operator $L_0$ we have chosen, we obtain precisely the expression \eqref{npeakon}.
We note that $u_x$ has jump discontinuities 
on the support of the singular measure $m$, so the meaning of $mu_x$ 
in the first equation \eqref{ch1} is {\it a priori\/} ambiguous.  
We shall show, however, that the derivative $D$ can be 
extended to piecewise smooth functions in such a way that all the operations 
extend to the discrete case without change. The operations 
which led {}from the overdetermined pair of equations \eqref{lxpair} to the nonlinear system \eqref{ch1} therefore continue to hold in the discrete case. In particular, the evolution of coupling coefficients \eqref{coupleev} applies to the multipeakon solutions as well. 

Suppose that a function $f$ has the form
\begin{equation}\label{piecewise} 
f(x)=f_j(x),\quad x_j<x<x_{j+1},
\end{equation}
where each $f_j$ belongs to $C^\infty(\mathbb R)$ and we have 
set $x_0=-\infty$,
$x_{n+1}=\infty$.  We normalize at the jumps by taking the average
of the limits {}from left and right:
$$
f(x_j)=\frac{f(x_j+)+f(x_j-)}{2}=\langle f(x_j)\rangle.
$$
If $g$ is a second such function normalized in the same way then the
distribution derivative $D$ satisfies the Leibniz rule: $D(fg)=fDg+gDf$.
Moreover, $D(fm)=fDm$.  

In the presence of smooth $t$ dependence,
\begin{equation}\label{tdepend} 
f(x,t)=f_j(x,t),\quad x_j(t)<x<x_{j+1}(t),
\end{equation}
$f$ as distribution has $t$-derivative 
\begin{equation}\label{tderiv}
\dot f =\frac{\partial f}{\partial t}-\sum^n_{j=1} [f(x_j+)-f(x_j-)]
\,\dot x_j \delta_{x_j}.
\end{equation}
Moreover
$$
\frac{d m}{dt} =\sum^n_{j=1}\big( \dot m_j\delta_{x_j}-m_j\dot x_j D\delta_{x_j}\big).
$$
In addition, we have
\begin{align*}
D(um)=&\frac{d}{dx}\sum_{j=1}^nu(x_j)m_j\delta_{x_j}=\sum_{j=1}^nu(x_j)
m_jD\delta_{x_j};\\[4mm]
u_x m=&\sum_{j=1}^n\langle u_x(x_j)\rangle m_j\delta_{x_j}.
\end{align*}

The first equation of \eqref{ch1} is therefore
$$
\sum_{j=1}^n\dot m_j\delta_{x_j}-m_j\dot x_jD\delta_{x_j}+\langle u_x(x_j)\rangle 
m_j\delta_{x_j}+u(x_j)m_jD\delta_{x_j}=0.
$$
Setting the coefficients of the independent distributions $\delta_{x_j}$ and 
$D\delta_{x_j}$ equal to zero, we obtain the Hamiltonian system
\begin{equation}
\dot x_j=\frac{\partial H}{\partial m_j}=u(x_j),
\qquad
\dot m_j=
-\frac{\partial H}{\partial x_j}=-\langle u_x(x_j)\rangle \,m_j\label{ev2}
\end{equation}
where
\begin{align}
H(x,m)=&\int^\infty_{-\infty}\Big(u^2+\frac14 u_x^2\Big)\,dx=\frac14\sum^n_{j,k=1}m_jm_k e^{-2|x_j-x_k|}\nonumber\\[4mm]
=&\frac12\sum^n_{j=1}u(x_j)\,m_j=\frac12 \int_{-\infty}^\infty u(x)\,dm(x).\label{hamiltonian}
\end{align}

\section{A Liouville transformation}\label{liouville}

The flow of the amplitudes $m_j$ and
the locations $x_j$ of the peaks can be computed explicitly, by virtue of the inverse scattering approach. As we shall see, there are $n$ eigenvalues $\lambda_j$
and coupling constants $c_j$.  The eigenvalues are fixed under
the evolution and the coupling constants evolve according to
\eqref{coupleev}.  In order to use this information to recover
the time dependence of $m_j$ and $x_j$ we find it convenient first to transform the
spectral problem.

In \cite{bss1} we obtained a Liouville transformation which transforms the spectral problem $L_0(z)\varphi=0$ to a ``density'' problem for a finite string, in which the density can take on both positive and negative values. 
The density problem with positive sign has been studied extensively by Krein 
\cite{krein1}, \cite{krein2}. In the research announcement \cite{bss2}
we used the classical results on continued fractions to obtain solutions of 
\eqref{chequ} in the case $m$ is a discrete 
measure, with weights of one sign.  The argument carries over to weights
of arbitrary sign.

For completeness, we repeat the details here.  The first step is to transform 
(a multiple of) the operator
$D^2-1$ that occurs in \eqref{lxpair} into the operator $D^2$ on 
the interval $(-1,1)$; then the wave functions for discrete $m$
become piecewise linear.  The appropriate coordinate transformation is
\begin{equation}\label{y}
y=\tanh\, x,\qquad \frac{dy}{dx}=\rho(x)=\frac1{\cosh^2 x}=1-y^2.
\end{equation}
The operator
\begin{eqnarray}\label{op1}
\frac1{\rho^2}\,L_0(z)={\cosh^4 x}\Big(\frac{d^2}{dx^2}-zm-1\Big)
\end{eqnarray}
is selfadjoint in $L^2(\R,\rho^2\,dx)$.  The operator
\begin{eqnarray}
U:L^2((-1,1),dy)\to L^2(\R,\rho^2\,dx),\label{unitary}\\[4mm]
[Uf](x)=\rho(x)^{-1/2}f(\tanh\, x)=(1-y^2)^{1/2}f(y)\nonumber
\end{eqnarray}
is unitary and carries \eqref{op1} to
\begin{eqnarray}\label{op3}
L(z)=\frac{d^2}{dy^2}-zg(y),\qquad g(y)=
m(\tanh^{-1} y)(1-y^2)^2=\frac{m(x)}{\rho(x)^2},
\end{eqnarray}
with Dirichlet boundary conditions.
As noted above, the resulting spectral problem
\begin{eqnarray}\label{gstring}
\frac{d^2v}{d y^2}(y)=z\,g(y)v(y),
\quad -1 <y<1;\qquad v(-1)=0=v(1),
\end{eqnarray}
is familiar when $g$ is positive: it determines the natural vibration
frequencies a string with mass density $g$.

The Liouville transformation takes the wave functions $\varphi_0$,
$\psi_0$ to the corresponding functions for the problem \eqref{gstring}:
\begin{eqnarray}
\varphi''-zg\varphi=0;\qquad \varphi(-1,z)=0,\quad \varphi'(-1,0)=1;\label{wave}\\[4mm]
\psi''-zg\psi=0;\qquad \psi(1,z)=0,\quad \psi'(1,z)=-1;\nonumber
\end{eqnarray}
here the primes denote derivatives with respect to $y\in[-1,1]$.
The eigenvalues $\{\lambda_\nu\}$ and coupling constants $\{c_\nu\}$
are the same as for the original problem. 

In particular, the evolution of the coupling coefficients is preserved
under the Liouville transformation.

The discrete measure \eqref{discrete1} is transformed into the discrete measure $g$
on the interval $(-1,1)$ given by 
\begin{equation}\label{discrete}
g=\sum^n_{j=1}g_j\,\delta_{y_j},\qquad -1=y_0<y_1<\dots<y_n<y_{n+1}=1.
\end{equation}
The terms in \eqref{discrete1} are related to the terms here  by \cite{bss2}
\begin{equation}\label{discrete2}
m_j=g_j(1-y_j^2),\quad x_j=\frac12\log\Big(\frac{1+y_j}{1-y_j}\Big).
\end{equation}

We remark that the Liouville transformation is a canonical transformation  
{}from the equations on the line to a Hamiltonian system on the interval
$(-1,1)$, but we shall not need that here. The multipeakon
solution is obtained by solving the inverse spectral problem for the
discrete string, and then pulling the solution back to the real line,
using \eqref{discrete2}.

In the next two sections we treat the direct and inverse spectral
problems for the discrete string. 

\section{The Weyl function}\label{weylfct}

Equation \eqref{gstring} implies that $\varphi$ is 
continous and piecewise linear, with jump discontinuities in the derivative 
at the points $\{y_j\}$.  
Then both sides of \eqref{gstring} are interpreted in the sense of 
distributions.  We denote left and right derivatives  
with respect to $y$ by $D_\pm$:
$$
D_+\varphi(y_j)=\frac{\varphi(y_{j+1})-\varphi(y_j)}{l_j},
\qquad
D_-\varphi(y_j)=\frac{\varphi(y_{j})-\varphi(y_{j-1})}
{l_{j-1}},
$$
where $l_j=y_{j+1}-y_{j}$, $0\le j\le n$.
Then \eqref{gstring} becomes
\begin{equation}\label{gstring2}
D_+\varphi(y_j)-D_-\varphi(y_j)=z\,g_j\,\varphi(y_j),\quad 1\le j\le n.
\end{equation} 

We note that the spectral problem may be written in the matrix form
\begin{equation}\label{jac}
Jq=zGq,
\end{equation}
where $q=(q_1,\dots,q_n)^t$, 
$q_j(t)=\varphi(y_j(t),t)$, 
$$
J=\begin{pmatrix} b_1       &    a_1   &    0     & \dots & 0    \\
        a_1  &    b_2        & a_2 &   0   &      \\
	0         &    a_2   &   b_3    &   &   \\
        \vdots    &                &         &   \ddots    &  a_{n-1}\\
        0         &               &          &  a_{n-1} & b_n
\end{pmatrix},
\qquad
G=\begin{pmatrix} g_1 & 0 & \dots & 0\\
0 & g_2 & \dots & 0 \\
\vdots &&& 0 \\
0  & \dots & 0 & g_n, 
\end{pmatrix}
$$
and $a_j=1/l_j$,  $b_j=-(a_{j-1}+a_j).$
The spectral problem for the multipeakon solutions
thus differs {}from that for the Jacobi flows, which simply takes the form 
$Jq=zq$.

The wave functions  $\varphi(\cdot, z)$ may be constructed recursively.  
We fix $z$ and set
$$
q_j=\varphi(y_j,z),
\qquad
 p_j=D_-\varphi(y_j),
$$
so that
\begin{equation}
 q_j-q_{j-1}=p_j l_{j-1}, \qquad
p_j-p_{j-1}=z\,g_{j-1}q_{j-1}.\label{pq}
\end{equation}
We begin with $q_0=0$ and $p_1=1$.
Then $q_j$ and $p_j$ are polynomials of degree $j-1$ in $z$. 
In particular, $\varphi(1,z)=q_{n+1}$ and $D_-\varphi(1,z)=p_{n+1}$
are polynomials in $z$ of degree $n$.
  
\begin{theorem}\label{roots} The set of roots of $\varphi(1,z)$ is the spectrum of the
problem \eqref{gstring}, equivalently \eqref{jac}.  The roots are real, 
simple, and non-zero.  The number of positive (resp.~negative) roots of
$\varphi(1,z)$ equals the number of negative (resp.~positive) terms
among the $g_j$.
\end{theorem}
{\em Proof:}  The eigenvalues are given by the zeroes of $\varphi(1,z)$.
Any other non-zero solution of \eqref{gstring} which vanishes at $-1$  
is a scalar multiple of $\varphi(y,z)$. Therefore there is at 
most one eigenfunction for each value of $z$, and the 
geometric multiplicity is one. Moreover, 0 cannot be an eigenvalue, 
since $\varphi(y,0)= 1+y$. The eigenvalues are real since 
$J$ and $G$ are real, symmetric matrices and $J$ is negative definite. 

Consider \eqref{jac} with $G=I$.  This is a pure eigenvalue
problem for the symmetric tridiagonal matrix $J$, so it has real spectrum.   
It corresponds to our spectral problem with all $g_j=1$; so if $z\ge 0$, then
the slope of $\varphi(\cdot,z)$ is non-decreasing {}from left to right and
necessarily $\varphi(1,z)>0$.  Consequently $J$ is negative
definite and has the form $-B^2$, where $B$ is positive definite.
Then \eqref{jac} is equivalent to
$B^{-1} G\,B^{-1} w=-z^{-1} w$, $w=Bq.$
Since $B^{-1} G B^{-1}$ is symmetric, the spectrum is real and the
algebraic multiplicity of each eigenvalue is its geometric multiplicity, 
which we have shown to be 1.

To obtain the conclusion about the signs of the roots, we set
$B(s)=(1-s)I+sB$, $0\le s\le 1$.  Each $B(s)$ is positive definite,
so by the previous argument the eigenvalues of
$B(s)^{-1} G B(s)^{-1}$ are simple and non-zero.  Therefore the number of
positive roots is independent of $s$.  Comparing $s=1$ and $s=0$, we
obtain the result. \quad{\it q.e.d.}

\medskip

We denote the roots of $\varphi(1,z)$ by $\{\lambda_j\}$.  Note that  
$\varphi(y,0)=1+y$, so $\varphi(1,0)=2$ and 
\begin{equation}\label{prod}
\varphi(1,z)=2\prod^n_{j=1}\Big(1-\frac{z}{\lambda_j}\Big).
\end{equation}
The {\it Weyl function\/} associated with \eqref{gstring2} is
\begin{equation}\label{weyl}
w(z)=\frac{D_-\varphi(1,z)}{\varphi(1,z)}=\frac{p_{n+1}}{q_{n+1}}.
\end{equation} 
It will be more convenient to work with the modified Weyl function
$w(z)/z$; in particular we use its partial fractions decomposition
\begin{equation}\label{modweyl}
\frac{w(z)}z=\frac1{2z}+\sum^n_{j=1}\frac{a_j}{z-\lambda_j}
=\sum_{j=0}^n\frac{a_j}{z-\lambda_j},
\end{equation}
where we have set
$$
a_0=\frac12,\quad \lambda_0=0.
$$

The {\it scattering data\/} for the spectral problem \eqref{gstring2}
consists of the spectrum $\{\lambda_j\}$ and the coupling
constants $\{c_j\}$.  Recall that the coupling constants relate
the eigenfunctions $\varphi$ to their counterparts $\psi$ normalized
at $y=1$ by $\psi(1,z)=0$, $D_-\psi(1,z)=-1$.  In fact
\begin{equation}
\varphi(y,\lambda_j)=c_j\psi(y,\lambda_j).\label{coupling}
\end{equation}

The Wronskian of two solutions of \eqref{gstring2} is constant in the 
intervals $y_j<y<y_{j+1}$ and continuous across the $y_j$. 
Evaluating the Wronskian $W(\varphi,\psi)$ at $y=\pm1$ 
establishes that $\varphi(1,z)=\psi(-1,z)$.  
Differentiating \eqref{coupling} with respect to $y$ and setting $y=1$, we obtain
\begin{equation}\label{coupling2}
c_j=-\varphi'(1,\lambda_j).
\end{equation}

The residues $a_j$, $j\ge 1$, in \eqref{modweyl} are determined by the 
scattering data.  Combining \eqref{coupling2} and \eqref{prod}, we obtain
\begin{equation}
a_j=\frac1{\lambda_j}\,
\frac{D_-\varphi(1,\lambda_j)}{\varphi_z(1,\lambda_j)}=
\frac{c_j}{2}\prod_{k\ne j}(1-\lambda_j/\lambda_k)^{-1}.\label{cjaj}
\end{equation}
Thus under the nonlinear evolution given by \eqref{chequ}, 
\begin{equation}\label{ajev}
\dot a_j=-\frac{2a_j}{ \lambda_j}, \qquad a_j(t)=a_j(0)e^{-2t/\lambda_j}.
\end{equation}

\begin{theorem}\label{aj} The residues $a_j$ in the partial fractions decomposition
\eqref{modweyl} of $w(z)/z$ are positive and satisfy
\begin{equation}
\lambda_j^2 \,\varphi^2_z(1,\lambda_j)\,
a_j=\int^1_{-1}\varphi^{\prime\,2} (y,\lambda_j)\,dy.\label{positive}
\end{equation}
\end{theorem}
{\em Proof:} Differentiation with respect to $z$ commutes with $D$, the
distribution derivative, so 
$$
D^2\varphi=z\varphi\,g,\qquad D^2\varphi_z=z\varphi_z\,g+\varphi\,g.
$$
We multiply the first of these equations by $\varphi_z$ and the
second by $\varphi$, subtract, and integrate with respect to $y$.
When $z=\lambda_j$ we obtain
\begin{equation}
\varphi'(1,\lambda_j)\,\varphi_z(1,\lambda_j)=
-\int^1_{-1}\varphi^2(y,\lambda_j)\,dg(y),
\label{pos1}
\end{equation}
where $dg$ is the measure defined by the distribution $g$ defined by \eqref{discrete}.
Integrating the equation $\varphi D^2\varphi =\lambda_j\varphi^2 g$ with
respect to $y$, we obtain
\begin{equation}
\int^1_{-1}\varphi^{\prime\,2}(y,\lambda_j)\,dy=-\lambda_j
\int^1_{-1}\varphi^2(y,\lambda_j)dg(y).\label{pos2}
\end{equation}
Combining \eqref{cjaj}, \eqref{pos1}, and \eqref{pos2}, we
obtain \eqref{positive}.\quad {\it q.e.d.}

\section{A theorem of Stieltjes}\label{stieltjes}

For the inverse spectral problem, we assume that the 
eigenvalues $\{\lambda_j\}$ and the residues
$\{a_j\}$ are known, and seek to determine the constants $\{g_j\}$
and the points $\{y_j\}$; in place of the latter we may look for
the subinterval lengths $l_j=y_{j+1}-y_j$.  A first step in the
process is to determine the Laurent expansion at infinity of the
modified Weyl function.  This is 
easily obtained {}from \eqref{modweyl} by expanding
$$
\frac{a_j}{z-\lambda_j}=\sum_{k=0}^\infty \frac{a_j\,\lambda_j^k}{z^{k+1}}.
$$
The result is the following.

\begin{lemma}\label{laurent} The modified Weyl function has the Laurent expansion
\begin{equation}\label{infty}
\frac{w(z)}{z}=\sum_{k=0}^\infty \frac{(-1)^kA_k}{z^{k+1}},
\end{equation}
where
\begin{equation}\label{Ak}
A_k=\sum_{j=0}^n(-\lambda_j)^ka_j.
\end{equation}
\end{lemma}

The Weyl function itself can also be written as a 
continued fraction, \cite{krein1}.
\begin{lemma}\label{weylfrac} The Weyl function is
\begin{equation}\label{confrac}
w(z)=\cfrac{1}{l_n+\cfrac{1}{zg_n+\cfrac{1}{l_{n-1}+\dots+\cfrac{1}
{zg_1+\cfrac{1}{l_0}  }}}}.
\end{equation}
\end{lemma}
{\em Proof:} We use \eqref{pq} inductively:
$$
p_1=1, \qquad q_1=l_0, \qquad q_2=q_1+l_1p_2, \qquad p_2=p_1+zg_1 l_0;
$$
\begin{equation}\nonumber
\frac{p_2}{q_2}=\frac{p_2}{l_0+p_2 l_1}=\dots 
=\cfrac{1}{l_1+\cfrac{1}{zg_1+\cfrac{1}{l_0}}}.
\end{equation}
Assuming \eqref{confrac} for $\{y_j\}_{j<n}$ and $\{g_j\}_{j<n}$,
we adjoin $y_n>y_{n-1}$ and $g_n$.   Then $p_{n+1}=p_n+zg_nq_n$, 
$q_{n+1}=q_n+l_np_{n+1}$; and
\begin{align}
\frac{p_{n+1}}{q_{n+1}}=&\frac{  p_{n+1}  }{   q_n+l_np_{n+1}  }=
\cfrac{1}{ l_n+\cfrac{q_n}{p_{n+1} }   }
=\cfrac{1} {l_n+\cfrac{q_n}{p_n+zg_nq_n}  }  \nonumber   \\[4mm]
=&\cfrac{1}{l_n+\cfrac{1}{zg_n+\cfrac{p_n}{q_n}  }  } \label{induction}
\end{align}
Hence \eqref{confrac} follows by induction.\quad {\it q.e.d.}

Dividing by $z$ gives the continued fraction decomposition of
the modified Weyl function:
\begin{equation}\label{modconfrac}
\frac{w(z)}z=
\cfrac{1}{l_n\,z+\cfrac{1}{g_n+\cfrac{1}{l_{n-1}\,z+\dots+\cfrac{1}
{g_1+\cfrac{1}{l_0\,z}  }}}}.
\end{equation}

\medskip
A classical result of Stieltjes \cite{stieltjes2}
recovers the coefficients of the
continued fraction \eqref{confrac} {}from the Laurent expansion of $w(z)/z$ at
infinity.  

A key role is played by the Hankel matrix whose entries are the
quantities $A_k$ defined by \eqref{Ak}. In the original theory of 
Stieltjes, the $A_k$ are all positive. This corresponds to the pure
peakon case, in which all the weights $g_j$ are positive and the
eigenvalues $\lambda_j$ are negative. In the peakon-antipeakon case
the weights are of both signs, and the $A_k$ are now of both signs; 
nevertheless,  --- and this will be an important point in the 
analysis ---  they are still the moments of a positive measure.
In fact \eqref{Ak} may be rewritten as
\begin{equation}\label{measure}
A_k=\int_{-\infty}^infty \lambda^kd\mu(\lambda),
\qquad
\mu=\sum^n_{j=0}a_j\delta_{-\lambda_j},
\end{equation}
where again we take $\lambda_0=0$ and $a_0=1/2$. 

In view of \eqref{Ak} and \eqref{coupling}, we see that the $A_j$ are
rational functions of the scattering data.  The following result
provides explicit formulas.

\begin{theorem}\label{stiel} {\rm [Stieltjes]} The Laurent series \eqref{infty}
can be uniquely developed in a continued fraction
$$
\cfrac{1}{b_1z+ \cfrac{1}{b_2+\cfrac{1}{b_3z+\dots}}},
$$
where
\begin{equation}\label{coefficients}
b_{2k}=\frac{(\Delta_k^0)^2}{\Delta_k^1\Delta_{k-1}^1},
\qquad
b_{2k+1}=\frac{(\Delta_k^1)^2}{\Delta_k^0\Delta_{k+1}^0}.
\end{equation}
Moreover
\begin{equation}\label{summation}
b_1+b_3+\dots+b_{2k+1}=\frac{\Delta_k^2}{\Delta_{k+1}^0}.
\end{equation}
Here the $\Delta^1_k$ are certain $k\times k$ 
minors of the infinite Hankel matrix
$$
H=\begin{pmatrix} A_0 & A_1 & A_2 & A_3 & \dots \\
A_1 & A_2 & A_3 & A_4 & \dots \\
A_2 & A_3 & A_4 & A_5 & \dots \\
A_3 & A_4 & A_5 & A_6 & \dots \\
\vdots & & & & \end{pmatrix}:
$$
$\Delta_k^l$ is the determinant of the $k\times k$ submatrix
 of $H$ whose (i,j) entry is $A_{l+i+j-2}$.  By convention,
$\Delta^0_l=1$.
\end{theorem}

Equations \eqref{coefficients} and \eqref{summation} appear in 
\cite{stieltjes2}, equations (7) and (11), respectively.
By comparing the continued fraction in \eqref{modconfrac} with that in 
Theorem \ref{stiel}, we obtain
\begin{equation}\label{lj}
l_j=\frac{(\Delta_{n-j}^1)^2}{\Delta_{n-j}^0\Delta_{n-j+1}^0}, \qquad
0\le j\le n; 
\end{equation}
\begin{equation}\label{gj}
g_j=\frac{(\Delta_{n-j+1}^0)^2}{\Delta_{n-j+1}^1\Delta_{n-j}^1}, 
\qquad 1\le j\le n;
\end{equation}
while {}from \eqref{summation} and \eqref{lj} we obtain
\begin{equation}\label{yj}
y_j=1-(l_n+l_{n-1}+\dots+l_j)=1-\frac{\Delta^2_{n-j}}{\Delta^0_{n-j+1}}.
\end{equation}

These results allow us to characterize the range of the forward spectral
map in terms of conditions on the $A_k$. 

\begin{lemma}\label{posdef} For even $l$, the minors $\Delta^l_k$ are strictly positive, $0\le k\le n$, as is $\Delta^0_{n+1}$.  
For odd $l$, and $0\le k\le n$, if all the weights $g_j$ have the same sign, 
then  $\Delta^l_k$ is non-zero and has the opposite sign, while  $\Delta^l_{n+1}=0$.
\end{lemma}
{\em Proof:}  The quadratic form 
associated with the $k\times k$ submatrix with upper left hand 
element $A_l$ is

$$
\sum_{i,j=0}^{k-1} A_{l+i+j}\xi_i\xi_j=\int_{-\infty}^infty 
\sum_{i,j=0}^{k-1} \xi_i\xi_j\lambda^{i+j+l}d\mu(\lambda)
=\int_{-\infty}^infty \lambda^l \xi^2(\lambda)\, d\mu(\lambda),
$$
where
$$
\xi(\lambda)=\sum_{j=0}^{k-1} \xi_j\lambda^j.
$$
Thus the form is positive definite when $l$ is even and $k\le n$:
a non-zero polynomial of degree $k-1$ cannot vanish at the
$n$ non-zero points $-\lambda_j$ in the support of $d\mu$.  (When $l=0$
the point $\lambda=0$ must also be considered.)
If all these
points $-\lambda_j$ are positive (respectively negative) then the
form is also positive (respectively negative) for odd $l$.
The associated determinant is $\Delta^l_k$.  

Finally, for odd $l$ and $k>n$, there is a nonzero polynomial 
$\sum c_j\lambda_{j}$
of degree $k-1$ that vanishes at each $\lambda_1$,\dots,$\lambda_n$, 
and it follows that the vector $(c_0,\dots,c_{k-1})^t$ is in the null space
of the matrix associated with $\Delta^l_{k}$. \quad {\it q.e.d.}

\smallskip

In the pure peakon or antipeakon case, {\it i.e.\/} when 
all the weights $m_j$ have the same sign, Theorem \ref{roots} 
says that all eigenvalues $\lambda_j$ have the same sign.
It follows immediately {}from Lemma \ref{posdef} that 
the minors $\Delta_j^1$ cannot vanish; hence $l_j>0$ 
for all time, and there are no exact collisions: the distances
$x_{k+1}-x_k$ are always strictly positive.

\begin{theorem}\label{range}  The real non-zero 
constants $\{\lambda_j,c_j\}$, $j=1,\dots n$
are scattering data for \eqref{gstring}
if and only if the $\lambda_j$ are distinct,
the $a_j$ given by \eqref{cjaj} are positive, and the moment
matrix constructed from the associated measure has the property
that the determinants $\Delta^1_j$, $0\le j\le n$, do not vanish.
\end{theorem}
{\em Proof:} Suppose that the $\{\lambda_j\}$, $\{c_j\}$ are scattering data.
We showed in \S\ref{weylfct} that the eigenvalues are distinct
and the $a_j$ are positive.  It follows {}from Lemma \ref{posdef} that
$\Delta^0_j>0$, $0\le j\le n+1$.
Since the $l_j$ are positive, it follows {}from \eqref{lj} that 
$\Delta^1_j\ne 0$, $0\le j\le n$.  

Conversely, starting from distinct $\{\lambda_j\}$
and $\{c_j\}$ such that $\{\alpha_j\}$ are positive and $\Delta^1_j\ne 0$,
\eqref{lj} and \eqref{gj} can be used to define $\{l_j\}$ and $\{g_j\}$. 
The Laurent series \eqref{infty} corresponds to the continued
fraction \eqref{confrac}.  The proof of Lemma \eqref{weylfrac} is reversible
The associated continued fraction \eqref{confrac} 
terminates, because $\Delta^1_{n+1}=0$,
and gives the Weyl function for the spectral problem with constants 
$\{g_j\}$ and $\{l_j\}$. \quad{\it q.e.d.}

To relate the measure $m$ to the scattering data, we begin with some 
notation and an identity.  
We denote by $\widetilde A_k$ and $\widetilde\Delta^1_k$ the 
corresponding moments and determinants with respect to the modified measure 
\begin{equation}\label{tildemu}
\widetilde \mu=\sum^n_{j=1}a_j\delta_{-\lambda_j}.
\end{equation}
These coincide with the $A_k$ and $\Delta^l_k$ except that $A_0=1/2+\widetilde A_0$,
and consequently 
\begin{equation}\label{wtdelta}
\Delta^0_k=\widetilde\Delta^0_{k}+\frac12\,\Delta^2_{k-1}, \qquad k\ge 1.
\end{equation}
It follows {}from \eqref{yj} and \eqref{wtdelta} that
\begin{equation}
1-y_j=\frac{\Delta^2_{n-j}}{\Delta^0_{n-j+1}},
\qquad
1+y_j=\frac{2\widetilde\Delta^0_{n-j+1}}{\Delta^0_{n-j+1}}.\label{oneplus}
\end{equation}

Combining \eqref{oneplus} with \eqref{gj} and
\eqref{discrete2} we obtain the following.

\begin{theorem}\label{line} The weights $m_j$ and positions $x_j$ associated with the distribution $m$ in \eqref{discrete1} are given by
\begin{equation}
x_j=\frac12\log\Big(\frac{2\widetilde\Delta^0_{n-j+1}}{\Delta^2_{n-j}}\Big);\qquad
m_j=\frac{2\widetilde\Delta^0_{n-j+1}\,\Delta^2_{n-j}}
{\Delta^1_{n-j+1}\,\Delta^1_{n-j}}.\label{mj}
\end{equation}
\end{theorem}

We illustrate with the cases $n=1,2,3$.  For $n=1$ we have
\begin{align*}
x_1=&\frac12\log\Big(\frac{2\widetilde\Delta^0_1}{\Delta^2_0}\Big)=
\frac12\log 2a_1;\\[4mm]
m_1=&\frac{2\widetilde\Delta^0_1\,\Delta^1_0}
{\Delta^1_1\,\Delta^1_0}=\frac{2a_1}{-\lambda_1a_1}=-\frac2{\lambda_1}.
\end{align*}
For $n=2$ the result is
\begin{align*}
x_1=&\frac12\log\frac{2(\lambda_1-\lambda_2)^2a_1a_2}{\lambda_1^2a_1+\lambda_2^2a_2},
\quad x_2=\frac12\log2(a_1+a_2);\\[4mm]
m_1=&-\frac{2(\lambda_1^2a_1+\lambda_2^2a_2)}
{\lambda_1\lambda_2(\lambda_1a_1+\lambda_2a_2)},
\quad m_2=-\frac{2(a_1+a_2)}{\lambda_1a_1+\lambda_2a_2}.
\end{align*}
Finally, for $n=3$ we have
\begin{gather*}
x_1=\frac12\log\frac{2(\lambda_1-\lambda_2)^2(\lambda_2-\lambda_3)^2(\lambda_3-\lambda_1)^2a_1a_2a_3}
{\sum_{j<k}\lambda_j^2\lambda_k^2(\lambda_j-\lambda_k)^2a_ja_k},\\[4mm]
x_2=\frac12\log\big(\frac{2\sum_{j<k}(\lambda_j-\lambda_k)^2a_ja_k}
{\sum\lambda_j^2a_j}\big),\quad x_3=\frac12\log\big(2\sum a_j\big);
\end{gather*}
\begin{gather*}
m_1=-\frac{2\sum_{j<k}\lambda_j^2\lambda_k^2(\lambda_j-\lambda_k)^2a_ja_k}
{\lambda_1\lambda_2\lambda_3\sum_{j<k}\lambda_j\lambda_k(\lambda_j-\lambda_k)^2a_ja_k},\\[4mm]
m_2=-\frac{2\sum \lambda_j^2a_j\sum_{j<k}(\lambda_j-\lambda_k)^2a_ja_k}
{\sum\lambda_ja_j \sum_{j<k}\lambda_j\lambda_k(\lambda_j-\lambda_k)^2a_ja_k},
\qquad
m_3=-\frac{2\sum a_j}{\sum\lambda_j a_j}.
\end{gather*}

Moser \cite{moser2} applied the theory of continued fractions for Jacobi 
matrices and obtained explicit solutions similar to these for the isospectral 
flow of $3\times 3$ Jacobi matrices in a special case.

\section{Asymptotics}\label{asymp}

The function $u$ in \eqref{npeakon} is a superposition of single peakons
($m_k>0$) and antipeakons ($m_k<0$) with peaks and troughs at the
$x_k$; however both the $m_k$ and the $x_k$ vary in time.  We show
now that $u$ is asymptotic at large times 
to superpositions of non-interacting
peakons and antipeakons with constant heights/depths:  
At large negative time peakons are found to the far left, decreasing in
height {}from left to right, travelling to the right at speeds proportional to
their heights, with antipeakons to the far right, increasing in depth
{}from left to right, travelling to the left at speeds proportional to
their depths.  At large positive times
antipeakons are to the left, with decreasing depths, and peakons to the
right, with increasing heights.  The difference, relative to a pure
superposition of such travelling solutions, is that each peak or
trough has undergone a phase shift.  The total (asymptotic) phase
shift for each peak or trough is precisely the sum of the phase
shifts that would occur in interactions with each of the other
terms separately (in a calculation with $n=2$); see \eqref{shift}. 

The key to the analysis of the long-term asymptotics is the evaluation of the determinants $\Delta^l_k$ and $\widetilde\Delta^0_k$.

\begin{theorem}\label{det} The determinants $\Delta^l_k$ of the moment matrix
for the measure $d\mu$ in \eqref{measure} are given by
\begin{equation}\label{delta}
\Delta^l_k=\sum_{J\subset\{0,1,\dots,n\}, |J|=k}a^J (-\lambda)^{lJ} \pi_J,
\end{equation}
where
$$
a^J=\prod_{j\in J}a_j,\quad \lambda^J=\prod_{j\in J}\lambda_j,
\quad \pi_J=\prod\limits_{j,m\in J, j<m}(\lambda_j-\lambda_m)^2.
$$
The determinants $\widetilde\Delta^0_k$ are
\begin{equation}\label{deltawt}
\widetilde\Delta^0_k=\sum\limits_{J\subset\{1,\dots,n\},\ |J|=k}a^J  \pi_J.
\end{equation}
\end{theorem}

The proof of this theorem is given at the end of this section.  We begin by 
using it to deduce the asymptotics,
We number the eigenvalues $\lambda_j$ so that
\begin{equation}\label{order}
\lambda_{m+1}<\dots<\lambda_n<\lambda_0=0<\lambda_1<\lambda_2<\dots<\lambda_m.
\end{equation}
An examination of the formulas \eqref{delta} and \eqref{wtdelta}
in the light of \eqref{ajev} and the assumption \eqref{order}
yields the following asymptotics of the determinants.

\begin{lemma}\label{asymptminus} As $t\to-\infty$, if $l>0$ and $k\le n$, then
$$
\Delta^l_k \sim\big(\prod_{j=1}^k a_j(0)(-\lambda_j)^l\big)\big(
\prod_{1\le j<r\le k}
(\lambda_j-\lambda_r)^2\big)\exp\Big(
-\sum_{j=1}^k\frac{2t}{\lambda_j}\Big).
$$
If $k\le m$ then
$$
\Delta^0_k \sim\big(\prod_{j=1}^k a_j(0)\big)\big(
\prod_{1\le j<r\le k}
(\lambda_j-\lambda_r)^2\big)
\exp\Big(-\sum_{j=1}^k\frac{2t}{\lambda_j}\Big).
$$
If $m<k\le n+1$, then
$$
\Delta^0_k \sim\frac12\big(\prod_{j=1}^{k-1}\lambda_j^2 a_j(0)\big)\big(
\prod_{1\le j<r<k}
(\lambda_j-\lambda_r)^2\big)\exp\Big(-\sum_{j=1}^{k-1}\frac{2t}{\lambda_j}\Big).
$$
In all cases, for $k\le n$
$$
\widetilde\Delta^0_k \sim\big(\prod_{j=1}^k a_j(0)\big)\big(
\prod_{1\le j<r\le k}(\lambda_j-\lambda_r)^2
\big)\exp\Big(-\sum_{j=1}^k\frac{2t}{\lambda_j}\Big).
$$
\end{lemma}
{\em Proof:} We discuss only the first case; the remaining are treated similarly.
The sum in \eqref{delta} is taken over all subsets of integers $J$ of size $k$.
The exponential term that corresponds to any such subset is
$$
\exp\left(-\sum_{j\in J}\frac{2t}{\lambda_j}\right).
$$
Since $-t\to\infty$, the dominant term is that for which $J=\{1,\dots ,k\}$. 
Hence
\begin{align*}
\Delta^l_k(t)\sim &\prod_{j=1}^k a_j(t)(-\lambda_j)^l\prod_{s<m\le k}
(\lambda_s-\lambda_m)^2\\[4mm]
=&\prod_{j=1}^k a_j(0)(-\lambda_j)^l\prod_{s<m\le k}(\lambda_s-\lambda_m)^2
\exp\left(-\sum_{j=1}^k\frac{2t}{\lambda_j}\right),
\end{align*}
and the first result follows.

\begin{lemma}\label{asymptplus} As $t\to+\infty$, if $l>0$ then for $k\le n$
$$
\Delta^l_k \sim\big(\prod_{j>n-k} a_j(0)(-\lambda_j)^l\big)\big(
\prod_{n-k< j<r}
(\lambda_j-\lambda_r)^2\big)\exp\Big(-\sum_{j>n-k}\frac{2t}{\lambda_j}\Big).
$$
If $k\le n-m$ then
$$
\Delta^0_k \sim\big(\prod_{j>n-k}^n a_j(0)\big)\big(
\prod_{n-k< j<r}^n
(\lambda_j-\lambda_r)^2\big)\exp\Big(-\sum_{j>n-k}^n\frac{2t}{\lambda_j}\Big).
$$
If $n-m<k\le n+1$, then
$$
\Delta^0_k\sim\frac12\big(\prod^n_{j>n-k+1}\lambda_j^2 a_j(0)\big)\big(
\prod^n_{n-k+1<j<r}
(\lambda_j-\lambda_r)^2\big)\exp\Big(-\sum^n_{j>n-k+1}\frac{2t}{\lambda_j}\Big).
$$
In all cases, for $k\le n$
$$
\widetilde\Delta^0_k \sim\big(\prod^n_{j>n-k} a_j(0)\big)\big(
\prod^n_{n-k< j<r}(\lambda_j-\lambda_r)^2
\big)\exp\Big(-\sum^n_{j>n-k}\frac{2t}{\lambda_j}\Big).
$$
\end{lemma}
Combining these results with \eqref{line}, we obtain the asymptotics
of the problem on the line.

\begin{theorem}\label{lineasympt} As $t\to-\infty$,
\begin{align}
x_{n-j+1}(t)\sim &-\frac{t}{\lambda_j} +\frac12\log\left[2a_j(0)\prod_{k=1}^{j-1}
\Big(1-\frac{\lambda_j}{\lambda_k}\Big)^2\right];\label{xjminus}\\
m_j(t)\sim & -\frac2{\lambda_j}.
\label{mjminus}
\end{align}
As $t\to+\infty$,
\begin{align}
x_j(t)\sim &-\frac{t}{\lambda_j}+\frac12\log\left[2a_j(0)\prod_{k=j+1}^n
\Big(1-\frac{\lambda_j}{\lambda_k}\Big)^2\right];\label{xjplus}\\
m_{n-j+1}(t)\sim & -\frac2{\lambda_j}.\label{mjplus}
\end{align}
\end{theorem}

It follows that the solution $u(x,t)$ is asymptotically a sum of 
free peakons ($\lambda_j<0$) and antipeakons ($\lambda_j>0$).  The term with asymptotic velocity  $-2/\lambda_j$ undergoes a phase shift due to the
interactions in the amount
\begin{equation}
\sum_{k=j+1}^n\log\Big|1-\frac{\lambda_j}{\lambda_k}\Big|-
\sum_{k=1}^{j-1}\log\Big|1-\frac{\lambda_j}{\lambda_k}\Big|.\label{shift}
\end{equation}

\smallskip

\noindent{\it Proof of Theorem \ref{delta}.\/}  The proof is based on the
following lemma.

\begin{lemma}\label{mommatrix} Let $A_k$, $k=0,1,2,\dots$, be the moments of the
discrete measure
$\sum_{j=0}^n b_j\delta_{\lambda_j}$:
$$
A_k=\sum_{j=0}^n b_j \lambda_j^k,
$$
where the weights $b_j$ are non-zero, but need not be positive. Then
the determinant 
\begin{equation}\label{detmom1}
\Delta^0_{n+1}=\begin{vmatrix}A_0&A_1&\dots&A_n\\
A_1&A_2&\dots &A_{n+1}\\
\vdots & & &\\
A_n&A_{n+1}&\dots&A_{2n}\end{vmatrix}
\end{equation} 
is given by
\begin{equation}\label{detmom2}
\Delta^0_{n+1}=\prod_{j=0}^n b_j \prod_{k>j}^n(\lambda_k-\lambda_j)^2.
\end{equation}
\end{lemma}
{\em Proof:}  $\Delta^0_{n+1}$ is a polynomial in the $b_j$ and $\lambda_j$,
so it is enough to verify the identity (6.10) when all the $b_j$ are
positive.  Under this asssumption, $\Delta^0_{n+1}$ is a moment matrix
for a positive measure with weights at the $\lambda_j$.
By Lemma \ref{posdef}, $\Delta^0_{n+1}$ is positive if and only the
$\lambda_j$ are distinct.  Therefore each root $\lambda_j-\lambda_k$
is double, and $\Delta^0_{n+1}$ is divisible by $\prod_{j<k}
(\lambda_j-\lambda_k)^2$.  Comparing the total degree 
in the $\lambda_j$, we find
\begin{equation}\label{detmom3}
\Delta^0_{n+1}=c_{n+1}(b_0,\dots,b_n) \prod_{k>j}(\lambda_k-\lambda_j)^2.
\end{equation}
The coefficient is a polynomial of total degree $n+1$ in the
$b_j$ and vanishes if any $b_j=0$, hence has the form
$c_{n+1}\prod b_j$.
The constant $c_{n+1}$ may be determined inductively:
we take $\lambda_n=0$, so only the moment $A_0$
has a $b_n$ term and $\Delta^0_{n+1}$ is the product of $b_n$ and
the $n\times n$ minor $\Delta^2_n$.  The latter is $\Delta^0_n$ for the
measure 
$$
\sum^{n-1}_{j=0}\lambda_j^2b_j\delta_{\lambda_j}
$$
and it follows that $c_{n+1}=c_n=\dots=c_1=1$.\quad{\it q.e.d.}

\medskip

Note that any larger moment matrix for the measure $d\mu$ supported
at $n+1$ sites vanishes, {\it e.g.\/} by introducing additional sites with
weights $b_{n+k}=0$.  

The formula \eqref{delta} for $l=0$, $k=n+1$  coincides with 
\eqref{detmom2}, under the replacement $\lambda_j\mapsto -\lambda_j$.  
The remaining formulas may be derived {}from this one.
First, suppose that $l=0$ and $k<n$.  We express the determinant
$\Delta^0_k$ as a sum of terms indexed by the $J$'s; each term is itself
the determinant of the $k\times k$ matrix associated with
the measure \eqref{measure}.  This proves \eqref{delta}
when $l=0$.  

When $l>0$,  $\Delta^l_k$
coincides with a determinant of the form $\Delta^0_k$ 
taken with respect to the measure
$$
\sum^n_{j=0}(-\lambda_j)^la_j\,\delta_{-\lambda_j}.
$$ 
Therefore the remaining formulas of \eqref{delta} are
consequences of the formulas for $l=0$.\qquad{\it q.e.d.}

\section{Collisions and close encounters}\label{collisions}

The phase shift formula \eqref{shift} suggests that interactions occur in 
pairs.  This is borne out by an analysis of close encounters 
($x_{k+1}(t_0)-x_k(t_0)$ positive
but locally minimal) and collisions ($x_{k+1}(t_0)=x_k(t_0)$). 
When only peakons or only antipeakons are present,
collisions cannot occur: the determinants $\Delta^1_j$ do not vanish;
see Lemma \ref{posdef}.
The equation $\dot x_j=u(x_j)$ in \eqref{ev2} shows that an overtaking peakon must be higher, and that after the event, the higher peakon must be to the right; whereas an overtaking trough must be lower, and moving to the left.  

\begin{figure}[!ht]
\centerline{\epsfxsize=\linewidth \epsfbox{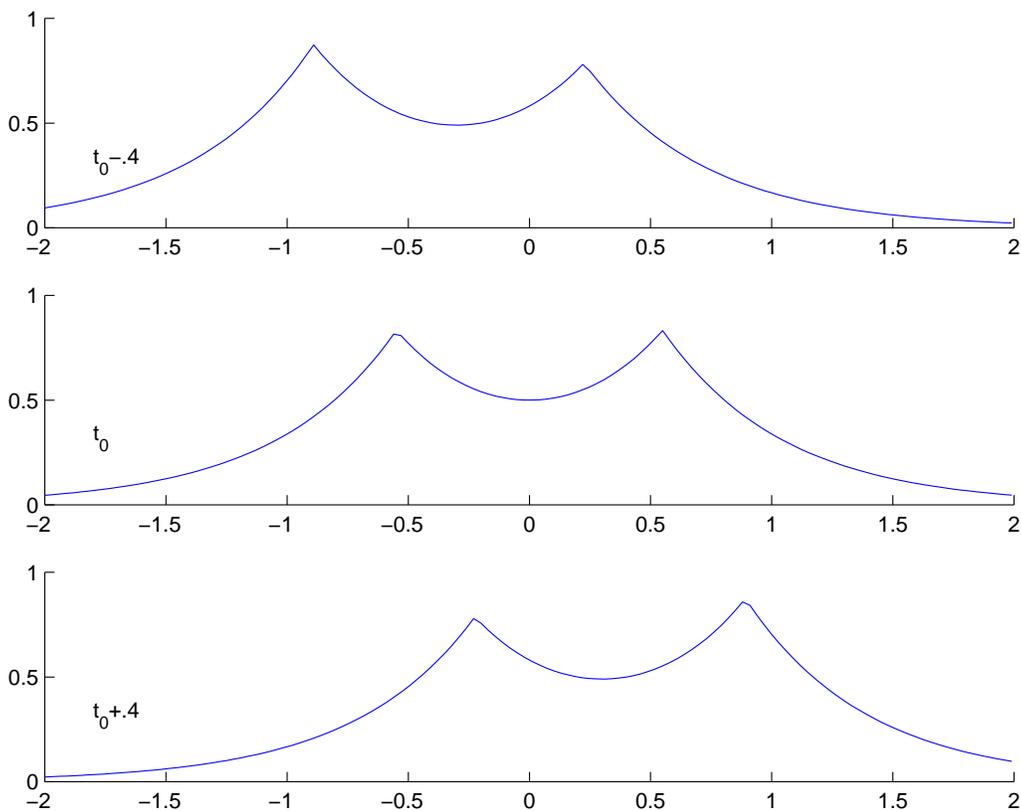}}
\caption{ A close encounter, computed {}from the exact formulas; $\lambda_1=-1$, $\lambda_2=-2$, $a_1(0)=1;\ a_2(0)=.5.$,
$t_0=0$. After the encounter, for $t>t_0$, the peakon on the right picks up speed and moves away.}
\end{figure}

When both peaks and troughs are present, collisions occur.
At a collision, some $\Delta^1_{n-k}=0$ and the solution of
the inverse problem for \eqref{gstring} breaks down:
the terms $m_k$ and $m_{k+1}$ blow up.  
We have $\Delta^1_0=1$, so $m_n$   
cannot become singular unless $m_{n-1}$ does also, and
\eqref{delta} implies that $\Delta_n^1>0$, so 
$m_1$ cannot become singular unless $m_2$ does also.

We show by a direct analysis of the inverse problem that
$u$ has a well-behaved continuation throughout the collision. (This can also be seen {}from the fact that the 
Hamiltonian \eqref{hamiltonian} is conserved under the flow, so that
the $W^{1,2}$ norm of $u$, and hence the $L^\infty$ norm, remain bounded.)   

The qualitative facts about collisions between peakons and 
antipeakons are summarized in the following theorem.

\begin{theorem}\label{limit} Suppose that some of the
determinants $\Delta^1_{n-k}$ vanish at $t=t_0$.  As $t$ approaches
$t_0$, collisions occur within distinct peakon-antipeakon pairs.
For $t<t_0$ each such pair consists of a peakon with
peak at $x_k$ and an antipeakon with trough at $x_{k+1}$; 
for $t_0<t$ the trough is at $x_k$ and the peak is at $x_{k+1}$.

As $t\to t_0\pm$ the function $u(x,t)$
converges uniformly (with respect to $x$) to a function 
that has the same form (but with the $x_j$ no longer distinct).
\end{theorem}

These aspects of the peakon-antipeakon collision are illustrated in the  
figure in the introduction.
\smallskip

Although $u(x,t)$ is well-behaved in the supremum norm at collisions,
this is not the case for the derivative.

\begin{theorem}\label{slope} Suppose that  $\Delta_{n-k}(t_0)=0$.  Then on the interval
$(x_k, x_{k+1})$,
\begin{equation}\label{uslope}
u_x(x,t)=\frac{\alpha}{t-t_0}+O(1)\quad {\rm as}\ \ t\to t_0,\qquad 
\alpha>0.
\end{equation}
\end{theorem}

The proofs of Theorems \ref{limit} and \ref{slope}
use some facts {}from the classical moment 
problem; proofs may be found in the monograph by Akhiezer, \cite{akhiezer}, 
Chapter 1.  Given the positive measure $d\mu$ of \eqref{measure},
one obtains  a sequence of $n+1$ polynomials $\{P_j(\lambda),\ 0\le j\le n\}$,
such that $P_j$ has degree $j$ and the $P_j$ 
are orthonormal in $L^2(\mathbb R,d\mu)$.  Explicitly  
\begin{equation}\label{orthopol}
P_j(\lambda)=\frac1{\big({\Delta^0_{j+1}\Delta^0_j}\big)^{1/2}}
\begin{vmatrix} A_0&A_1&\dots&A_j\\
A_1&A_2&\dots& A_{j+1}\\
\vdots&&\\
A_{j-1}&A_{j}&\dots& A_{2j-1}\\
1&\lambda&\dots&\lambda^j\end{vmatrix},\qquad 0\le j\le n.
\end{equation}
In particular,
\begin{equation}
P_j(0)=(-1)^j\frac{\Delta^1_j}{\big({\Delta^0_{j+1}
\Delta^0_j}\big)^{1/2}},\quad 0\le j\le n.
\label{pj0}
\end{equation}
hence, by \eqref{lj}, a collision between the $k^{th}$ and $(k+1)^{st}$
places occurs precisely when the constant term in the polynomial 
$P_{n-k}(\lambda)$ vanishes. 

The $P_j(\lambda)$ satisfy a second order recursion relation
\begin{equation}\label{recursion}
\lambda P_j(\lambda)=b_jP_{j+1}(\lambda)+d_jP_j(\lambda)+
b_{j-1}P_{j-1}(\lambda),\quad 1\le j\le n-1,
\end{equation}
where
$$
b_j=\frac{\big({\Delta^0_{j}
\Delta^0_{j+2}}\big)^{1/2}  }  {\Delta^0_{j+1}}>0,
\quad 0\le j\le n-1.
\label{bk}
$$
The recursion relation implies a well-known formula of
Darboux-Christoffel:
$$
b_j\,\frac{P_{j+1}(\lambda)P_j(\lambda')-P_j(\lambda)P_{j+1}(\lambda')}
{\lambda-\lambda'}=\sum_{i=0}^j P_i(\lambda)P_i(\lambda'),\quad 0\le
j\le n-1.
$$
We shall use the limiting form
\begin{equation}\label{darboux}
b_j\big(P'_{j+1} P_j-P_{j+1} P'_j\big)=\sum^j_{i=0}P_i^2,\quad
0\le j\le n-1.
\end{equation}
where primes denote differentiation with respect to $\lambda$.

\begin{theorem}\label{momform} The multipeakon solutions may be expressed in terms of the orthogonal polynomials $P_j(0,t)$ as follows:
\begin{gather}
l_j=P_{n-j}(0)^2,\quad g_j=-\frac1{b_{n-j}P_{n-j+1}(0)\,P_{n-j}(0)}\label{lgbp}
\\[4mm]
y_j=1+
b_{n-j}\Big(P'_{n-j}(0)P_{n-j+1}(0)-P_{n-j}(0)P'_{n-j+1}(0)\Big).\label{yjmom}
\end{gather}
\end{theorem}
{\em Proof:} The formulas follow immediately {}from \eqref{lj}, \eqref{gj},
\eqref{pj0} and \eqref{darboux}.

\medskip

An immediate consequence of \eqref{darboux} and the assumption that 
$P_0(0)\ne 0$ is that no two consecutive
$P_j(0)$ can vanish simultaneously.  Therefore no two consecutive $\Delta^1_j$
can vanish, $0\le j\le n-1$; and collisions can occur only in
distinct pairs $m_k$, $m_{k+1}$.  Moreover, if $P_j(0)=0$ then
$P'_j(0)\ne 0$.  {}From now on we consider the time-dependent case,
but we shall mainly suppress the time dependence in the notation, 
{\it e.g.} $P_i=P_i(\lambda,t)$, $P_i(0)=P_i(0,t)$.

\begin{lemma}\label{delta1} Suppose that the measure \eqref{measure} evolves according
to \eqref{ajev},
and let $\{P_j(\lambda,t)\}$ be the 
associated orthogonal polynomials.  Suppose that  $P_j(0,t_0)=0$.  Then the
derivative $\dot P_j$ at $t=t_0$ satisfies
\begin{equation}\label{pjdot}
\dot P_j+2P_j/\lambda
=P'_j(0)\sum_{i<j}P_i(0)\,P_i.
\end{equation}
\end{lemma}
{\em Proof:} Let $(\ ,\,)$ denote the inner product with respect to
$d\mu$.  Differentiating $(P_j,P_i)=\delta_{ij}$, 
with respect to $t$ (and suppressing the time variable) gives
\begin{equation}\label{pjdot2}
0=(\dot P_j,P_i)+(P_j,\dot P_i)+\sum_{\lambda_m\ne 0}\frac{2P_j(-\lambda_m)}
{-\lambda_m}\,P_i(-\lambda_m)\,a_m.
\end{equation}
We assume that $P_j(0)$ vanishes at $t=t_0$; then $P_j/\lambda$
is a polynomial of degree less than $j$ with constant term $P'_j(0)$
and we can add a summand at $\lambda=0$ in \eqref{pjdot2} to obtain
$$
(\dot P_j,P_i)+(P_j,\dot P_i)+2(P_j/\lambda,P_i)=P'_j(0)P_i(0).
$$
Note that $(P_j,\dot P_i)=0$ if $i<j$.
It follows that the polynomial $\dot P_j+2P_j/\lambda$ has degree $<j$
and its expansion in the orthonormal basis $\{P_i\}$ is
\eqref{pjdot}.
\qquad{\it q.e.d.}

\medskip

Note that \eqref{pjdot} implies 
\begin{equation}\label{pjdot3}
\dot P_j(0)+2P'_j(0)
=P'_j(0)\sum_{i<j}P_i(0)^2.
\end{equation} 

The next lemma implies that the zeros of the $\Delta^1_j(t)$ are simple.

\begin{lemma}\label{delta2} Under the assumptions of Lemma \ref{delta1}, 
$\dot P_j(0)P'_j(0)<0$ at $t=t_0$.
\end{lemma}
{\em Proof:}  Since the polynomial $\dot P_j+2P_j/\lambda$ has degree $<n$, it
cannot vanish at all the non-zero points $-\lambda_m$ in the support of
the measure.  Therefore we
have a strict inequality
\begin{equation}\label{pjdot4}
\frac12\big(\dot P_j(0)+2P'_j(0)\big)^2< ||\dot P_j+2\lambda^{-1} P_j||^2
 =\sum_{i<j}P'_j(0)^2 P_i(0)^2,
\end{equation}
where $||\ ||^2$ is the norm on $L^2(\R,d\mu)$. The equality in \eqref{pjdot4} follows {}from \eqref{pjdot} and the orthonormality of the polynomials $P_j$.

It follows {}from \eqref{pjdot3} and \eqref{pjdot4} that
$$
-2\dot P_j(0)P'_j(0)> \dot P_j(0)^2.\qquad{\it q.e.d.}
$$

\smallskip

The next lemma allows us to determine the direction of
the sign change for interacting peakon/antipeakon pairs.

\begin{lemma}\label{delta3} Under the assumptions of  Lemma \ref{delta1}, at $t=t_0$
\begin{equation}\label{pjdot5}
\frac{d}{dt}\big(P_{j+1}(0,t)P_j(0,t)\big)>0,\qquad
\frac{d}{dt}\big(P_{j}(0,t)P_{j-1}(0,t)\big)<0,
\end{equation}
for $1\le j\le n-1$ and $1\le j \le n$ respectively.
\end{lemma}
{\em Proof:} By Lemma \ref{delta2} and the assumption that $P_j(0,t_0)=0$,
the derivatives in \eqref{pjdot5} have the same sign as 
$$
-P_{j+1}(0,t_0)P'_j(0,t_0),\qquad -P'_j(0,t_0)P_{j-1}(0,t_0),
$$
respectively.  The conclusion \eqref{pjdot5} follows {}from
\eqref{darboux}.\quad{\it q.e.d.}

\medskip

The final lemma gives an algebraic proof that the singularities in successsive
$g_k$'s, hence in successive $m_k$'s, cancel each other
in a collision.

\begin{lemma}\label{delta4} Under the assumptions of Lemma \ref{delta1}
the sum $g_k+g_{k+1}$
has a finite limit as $t\to t_0$.
\end{lemma}
{\em Proof:} Let $j=n-k$.  {}From \eqref{lgbp} and \eqref{recursion} we obtain
$$
g_k+g_{k+1}=-\frac{b_jP_{j+1}(0)+b_{j-1}P_{j-1}(0)}
{b_j b_{j-1}P_{j+1}(0)P_j(0)P_{j-1}(0)}
=\frac{d_j}{b_jb_{j-1}P_{j+1}(0)P_{j-1}(0)}.
$$
The denominator does not vanish for $t$ near $t_0$, and it follows {}from \eqref{recursion} that $d_j=(\lambda P_j,P_j),$
which has a finite limit.   {\it q.e.d.}

\medskip

{\it Proof of Theorem \ref{limit}\/}.  As noted above, \eqref{darboux}
implies that collisions occur only in pairs $m_k$, $m_{k+1}$, where
$\Delta^1_{n-k}$ vanishes.  According to Theorem \ref{line} and
equation \eqref{pj0},  $m_k$ and $m_{k+1}$ have the same signs as
$-P_{j+1}(0)P_j(0)$ and $-P_j(0)P_{j-1}(0)$ respectively, $j=n-k$.
It follows {}from \eqref{pjdot5} that $m_k$ changes {}from positive to negative 
and $m_{k+1}$ changes {}from negative to positive.  In other words,
a peakon on the left changes places with an antipeakon on the right.

The assertion about uniform convergence of $u(x,t)$ follows {}from
Lemma \ref{delta4}.  In fact suppose that $\Delta^1_{n-k}(t_0)=0$, so
that $m_k$ and $m_{k+1}$ become singular.  At the same time,
however, $y_k-y_{k+1}\to 0$, so Lemma \ref{delta4} and \eqref{discrete2}
imply that the sum $m_k+m_{k+1}$ has a finite limit.  
For such a pair
\begin{multline*}
m_k(t)e^{-2|x-x_k(t)|}+m_{k+1}(t)e^{-2|x-x_{k+1}(t)|}\\[4mm]
=\{m_k(t)+m_{k+1}(t)\}e^{-2|x-x_k(t)|}
+ m_{k+1}(t)\{e^{-2|x-x_{k+1}(t)|}
-e^{-2|x-x_k(t)|}\}.
\end{multline*}
The difference between the exponentials is uniformly 
$O\big(|x_k(t)-x_{k+1}(t)|\big)$, so the sum converges uniformly to
$$
m_k(t_0)e^{-2|x-x_k(t_0)|}+m_{k+1}(t_0)e^{-2|x-x_{k+1}(t_0)|},
$$
where
\begin{align*}
m_k(t_0)=&m_{k+1}(t_0)=\frac12\,\lim_{t\to t_0}\{m_k(t)+m_{k+1}(t)\},
\\[4mm]
x_k(t_0)=&\lim_{t\to t_0}x_k(t)=\lim_{t\to t_0}x_{k+1}(t)=x_{k+1}(t_0).
\end{align*} 
The same argument applies to any other pair that become singular
at $t=t_0$, so $u(x,t)$ converges uniformly. \quad{\it q.e.d.}

\medskip

{\it Proof of Theorem \ref{slope}\/}.  It follows {}from the results of
this section that if $\Delta^1_{n-k}(t_0)=0$, then 
\begin{align}
x_{k+1}-x_k&=O(l_k)=\alpha_0(t-t_0)^2+O((t-t_0)^3);\nonumber\\
m_k&=-\frac{\alpha_1}{t-t_0}+O(1);\nonumber\\
m_{k+1}&=\frac{\alpha_1}{t-t_0}+O(1),\nonumber
\end{align}
for some positive constants $\alpha_j$ which can be computed explicitly.  
{}from these facts and \eqref{npeakon} we deduce that on the interval
in question we may compute the derivative to order $O(1)$ by
dropping all but two summands and replacing them by
$$
\frac{\alpha_1}{t-t_0}\big( e^{2x-2x_{k+1}}-e^{2x_k-2x}\big)
$$
whose derivative on the interval $(x_k,x_{k+1})$
is $4\alpha_1/(t-t_0)+O(1)$.\qquad{\it q.e.d.}

\end{document}